\documentclass[conference,usenatbib]{basi}
\usepackage[T1]{fontenc}
\usepackage[british]{babel}
\usepackage[varg]{txfonts}
\usepackage{rotating}
\usepackage{dcolumn}
\begin{document}
\title[Role of Viscosity and Cooling in Simulation of TCAF]{Role of Viscosity and Cooling in Hydrodynamic 
Simulation of Two Component Accretion Flow (TCAF) around Black Holes}
\author[Giri and Chakrabarti]%
       {Kinsuk Giri$^1$\thanks{email: \texttt{kinsuk@bose.res.in}}  and
       Sandip K. Chakrabarti$^{1,2}$\\
       $^1$ S.N. Bose National Centre for Basic Science, Salt Lake, Kolkata 700098, India \\
       $^2$ Indian Centre for Space Physics, Chalantika 43, Garia Station Rd., Kolkata 700084, India}

\pubyear{2013}
\volume{**}
\pagerange{**--**}
%\status{submitted}

\date{Received --- ; accepted ---}

\maketitle
%------------------------------------------------------------------------------%
% abstract and keywords                                                        %
%------------------------------------------------------------------------------%
\label{firstpage}

\begin{abstract}
We carry out numerical simulation of accretion flows around a black hole in presence of both viscous and cooling
effects. Instead of using a constant $\alpha$ parameter throughout the simulation grid,
we assume that $\alpha$ is maximum on the equatorial plane and gradually goes down
in the perpendicular direction We show that  when the injected sub-Keplerian flow
angular momentum is high enough and/or the viscosity and also cooling is high enough, Two Component
{\bf Accretion} Flow (TCAF) would be formed, otherwise the sub-Keplerian flow would
remain sub-Keplerian. We see that a Keplerian disk is produced on the equatorial plane.
Time variations of the total, Keplerian and Sub-Keplerian matter are studied with respect
to various flow parameters.
\\[6pt]
\hbox to 30pt{\hfil}\verb|http://www.ncra.tifr.res.in/~basi/|
\end{abstract}


\section{Introduction}
There are variety of astrophysical situations in which one expects to find matter accreting onto
a black hole. Most of the popular studies in accretion disks consider steady-state situations
\citep{B52,SS73,NT73,C89,P98} i.e. where things are not changing with time. The first numerical
attempt to study the behaviour of matter around black holes was made nearly three decades
ago \citep{HSW84}. \citep{CM93} presented results of numerical simulations
of thin accretion disks and winds. The most significant conclusion was that shocks in an
inviscid flow \citep{C89} were extremely stable . \citep{MSC94} extended their earlier numerical
simulation of accretion disks with shock waves when
cooling effect are also included. \citep[][Hereafter CT95]{CT95} proposed that the spectral
properties are better understood if the disk solutions of sub-Keplerian flows are included along with the
Keplerian flows which is known as two component accretion flows (TCAF). In TCAF,
the viscous Keplerian disk resides in the equatorial plane,
while the weakly viscous sub-Keplerian flow flanks the
Keplerian component both above and below the equatorial
plane. The two components merge into a single component
when the Keplerian disk also become sub-Keplerian.
In this paper, we have shown the effect of power law cooling as well as the turbulent viscosity
in the accretion flows towards a black hole  with time dependent simulation.
We found TCAF which was assumed in CT95 is obtained in our numerical simulation.  

\section{Model Equations and Methodology}
The basic equations describing a two-dimensional axisymmetric inviscid flow around a Schwarzschild 
black hole can be described as \citep{GCMR10}
The self-gravity of the accreting matter is ignored. Cylindrical coordinates 
$(r, \phi, z)$ is adopted with the z-axis being the rotation axis of the disk. 
The equations governing the viscous flow have been presented in \citep{GC12}
We have incorporated a cooling in the energy equation
in with a volumetric power law cooling rate ${{\Lambda}_{powcool}}  \propto {{\rho}^2}T^{\beta}$, where,
$\beta$ is the cooling index which is an arbitrary constant. 
In the present work, 
instead of using the constant viscous parameter $\alpha$ for the whole $x-z$ plane, we choose a smooth the distribution as,
$$
\alpha = {\alpha}_{max} - [{\alpha}_{max}{({z \over {r_{max}}})^{\delta}}],  \eqno(1)
$$
where, $r_{max} = 200, 0 \leq z \leq 200$ and $\delta > 0$. In our cases, we have chosen 
$\delta = 1.5$. If turbulence is the major source of viscosity, then, clearly it will be highest on the equatorial plane. 
We also assume that the gravitational field of the black hole can be described by \citep{PW80}.
All the simulations have been carried out assuming 
a stellar mass black hole $(M = 10{M_\odot})$. 
We carry out the simulations for several hundreds of dynamical 
time-scales. In reality, our simulation time corresponds to a few {\bf ten} seconds in physical units.

\section{Results and Discussions}
\begin{figure}
\begin{center}
{\includegraphics[width=4.5cm]{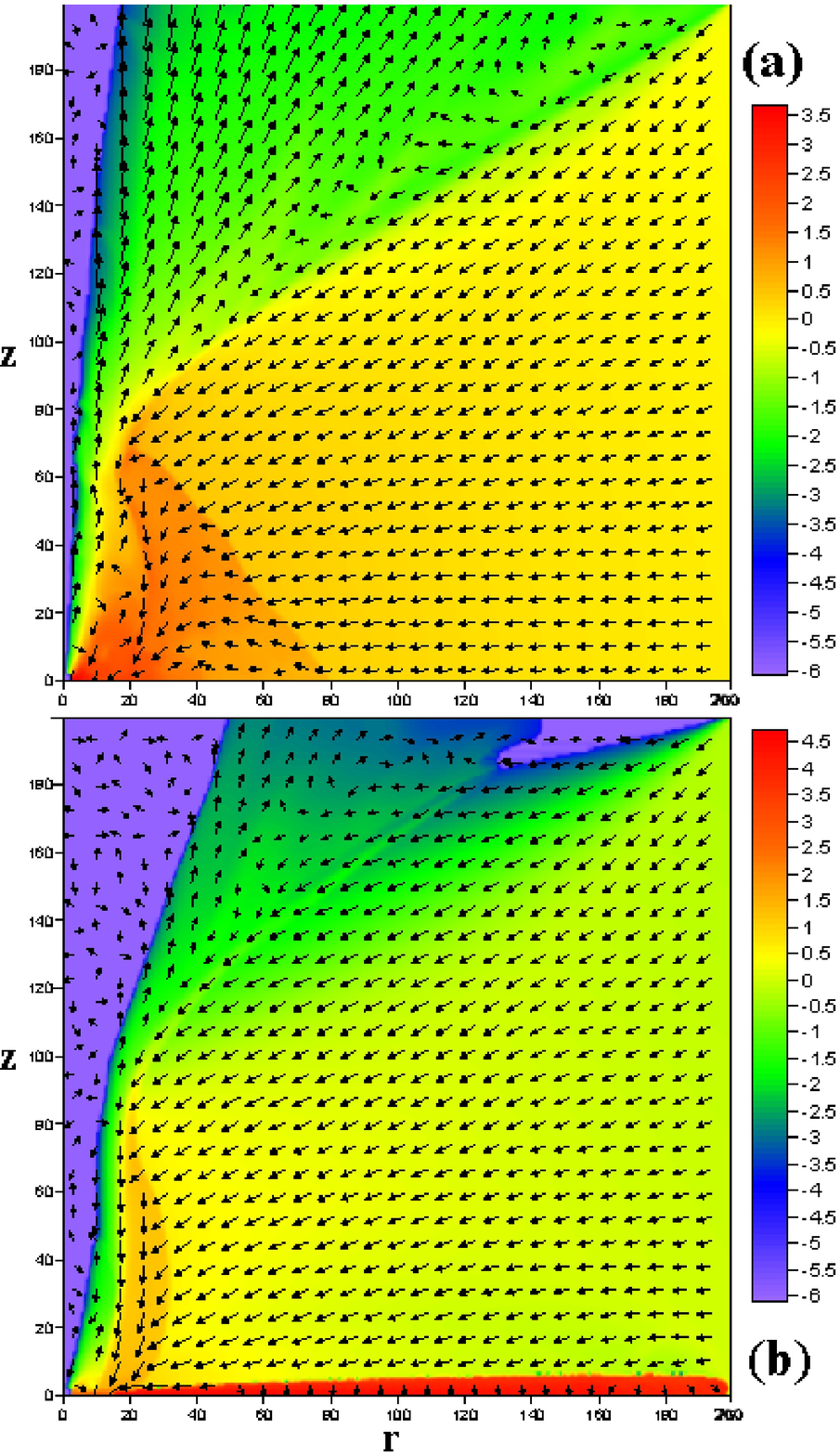}}
{\includegraphics[width=4.5cm]{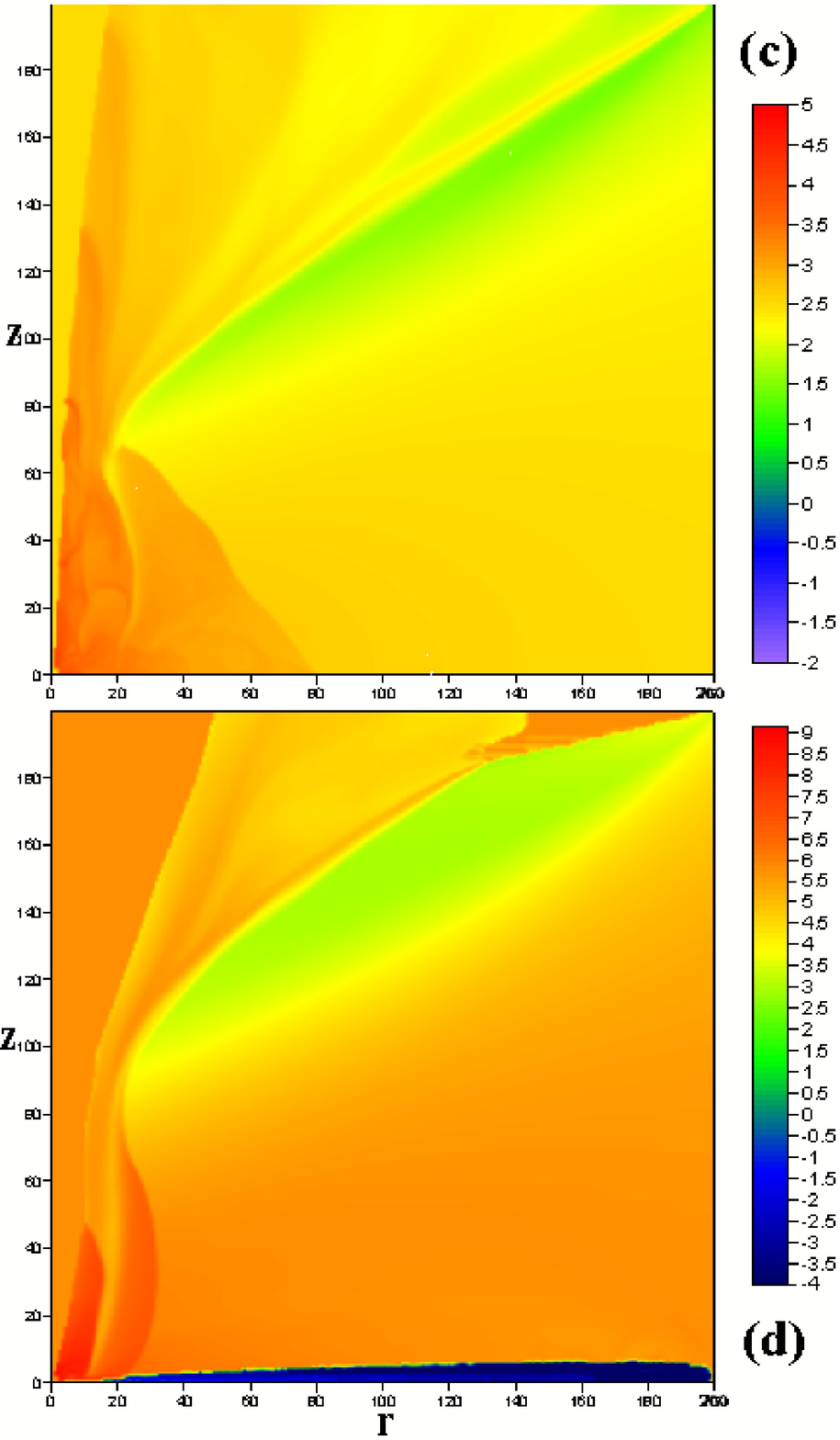}}
\caption{{\bf Distribution of  $(i)$ density and velocity}  $(ii)$ temperature distributions at $t= 95$s (a)/(c) without and (b)/(d) 
the inclusion of viscosity and cooling. Densities and temperatures in normalized units are plotted in logarithmic scale as in the scale on the right.
A two component flow is clearly formed in (b) and (d). {\bf The color scales for each distributions  are shown at the thin rectangular boxes }}
\end{center}
\end{figure}
We assume the flow to be in vertical equilibrium (C89) at the outer boundary.
The injection rate of the momentum density is kept uniform throughout the injected height at the outer edge.
We inject through all  grids. We stop the simulation at $t=94.86$ seconds. This is more than two hundred times the
dynamical time. Thus, the solution has most certainly come out of the transient 
regime and started exhibiting solutions of steady state.
In Fig. 1a, we show the velocity and density distribution of the flow without viscosity and cooling 
and in Fig. 1b we show flow with viscosity and cooling.
Here, for both the cases, {\bf the specific angular momentum $\lambda) = 1.7$ and 
specific energy ${\cal E}) = 0.001$}. We have taken ${\alpha}_{max} = 0.012$ and $\beta = 0.6$ for the later case.
All the density distributions are taken in logarithmic scale.
In Fig. 1(c-d), we show the temperature distributions in keV in logarithmic scale.
In the absence of cooling and viscosity, in Fig. 1a \& 1c, the single component sub-Keplerian flow forms. 
Physically, a viscous heating increase post-shock pressure and also transports
angular momentum faster. As a result the Rankine-Hugoniot condition is satisfied away
from the black hole. Cooling, on the other hand, reduces the post-shock pressure.
In Fig. 1b \& 1d, because of higher viscosity, flows have
the Keplerian distribution near the equatorial region. Because of cooling effects, the
region with a Keplerian distribution is cooler and denser. Comparatively low dense
sub-Keplerian matter stays away from the equatorial plane. For both the cases, the
Centrifugal Pressure supported BOundary Layer (CENBOL) forms.
Thus  the flows having sufficiently high viscosity on the equatorial plane and low viscosity above
and below, produces a TCAF where a Keplerian disk is 
surrounded by a rapidly moving sub-Keplerian halo. It is interesting to study the time variations of 
Keplerian matter (${M_{kep}}$), sub-Keplerian matter
(${M_{subkep}}$) and the total matter (${M_{tot}}$) in the system.
In Fig. 2, we have shown the time variations of total (${M_{tot}}$), Keplerian (${M_{kep}}$)
and sub-Keplerian (${M_{subkep}}$) matter
of the flow with viscosity and cooling. The solid curve represent the total matter of the system
with time, while dotted and dashed curve represent the variation of Keplerian and sub-Keplerian
matter respectively. Thus, we see that viscosity and cooling plays an important role in the time variations of
Keplerian and sub-Keplerian matters in the system.

\begin{figure}
\centerline{\includegraphics[width=5cm]{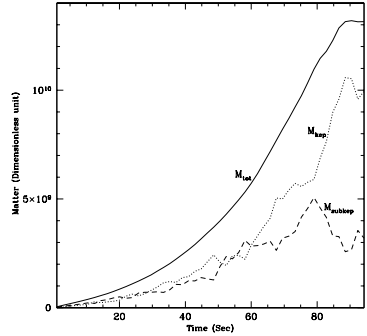}}
\caption{Time variations of the Total, Keplerian, sub-Keplerian
and out flowing matter in the system. See text for details.}
\end{figure}
\section*{Acknowledgments}
KG acknowledges the conference organizers (in particular to Dr. Santabrata Das) for the financial support to participate 
RETCO-2013.

%------------------------------------------------------------------------------%
% appendices:                                                                  %
%------------------------------------------------------------------------------%
%------------------------------------------------------------------------------%

\begin{thebibliography}{}

\bibitem[Bondi (1952)Bondi]{B52} Bondi, H., 1952, MNRAS, 112, 195
\bibitem[Chakrabarti (1989)Chakrabarti]{C89} Chakrabarti, S.~K., 1989, ApJ 347, 365 (C89)
\bibitem[Chakrabarti \& Molteni (1993)Chakrabarti \& Molteni]{CM93} Chakrabarti, S.~K., Molteni, 1993, ApJ, 417, 672
\bibitem[Chakrabarti \& Titarchuk (1995)Chakrabarti \& Titarchuk]{CT95} Chakrabarti, S.~K., Titarchuk, L.~G., 1995, ApJ, 455, 623 
\bibitem[Hawley et al. (1984)Hawley et al.]{HSW84} Hawley, J.~F., Smarr, L.~L., Wilson, J.~R., 1984, ApJ, 277, 296
\bibitem[Giri et al.(2010)Giri er al.]{GCMR10} Giri, K., Chakrabarti, S.~K., Samanta, M.~M.,  Ryu, D., 2010, MNRAS, 403,516
\bibitem[Giri \& Chakrabarti (2012)Giri \& Chakrabarti]{GC13} Giri, K., Chakrabarti, S. ~K., 2012, MNRAS, 421, 666
\bibitem[Molteni et al. (1996)Moltebi et al.]{MSC96} Molteni, D., Sponholz, H., Chakrabarti, S.~K., 1996, ApJ, 457, 805
%\bibitem{9} Narayan, R., Yi, I., 1994, ApJ, 428, L13
\bibitem[Novikov \& Thorne (1973)Novikov \& Thorne]{NT73} Novikov, I.~D., Thorne, K.~S., 1973, in Black Holes, ed. B. S. De Witt and C. De Witt (New York: Gordon \& Breach), 343
\bibitem[Paczy\'nski \& Witta (1980)Paczy\'nski \& Witta]{PW80} Paczy\'nski, B., Wiita, P. ~J., 1980, A \& A,  88, 23
\bibitem[Paczy\'nski (1998)Paczy\'nski]{P98} Paczy\'nski, B., 1998, Acta Astronomica, 48, 667
\bibitem[Shakura \& Sunyaev (1973)Shakura \& Sunyaev]{SS73} Shakura, N. ~I., Sunyaev, R. ~A., 1973, A \& A, 24, 337
\end{thebibliography}
\end{document}